\documentclass[useAMS,usenatbib,usegraphicx]{mn2e}

\usepackage{xspace}
\usepackage[colorlinks=true,citecolor=blue]{hyperref}
\usepackage{natbib}

\newcommand{\MJup}{M$_{\mathrm{Jup}}$\xspace}

\newcommand{\teff}{T$_{e\! f\! f}$\xspace}
\newcommand{\logg}{log~\emph{g}\xspace}

\newcommand{\mic}{$\umu$m\xspace}
\newcommand{\as}{\hbox{$^{\prime\prime}$}\xspace}
\newcommand{\am}{\hbox{$^{\prime}$}\xspace}
\newcommand{\degre}{$^\circ$\xspace}

\title[Exoplanet characterization using ADI and SDI]{Photometric characterization of exoplanets using angular and spectral differential imaging\thanks{This version of the paper is for astro-ph. The definitive version is available at \href{http://www.blackwell-synergy.com/}{http://www.blackwell-synergy.com/}.}}

\author[A. Vigan et al.]
{A. Vigan$^{1}$\thanks{E-mail: \href{mailto:arthur.vigan@oamp.fr}{arthur.vigan@oamp.fr}},
C. Moutou$^{1}$,
M. Langlois$^{2}$,
F. Allard$^{3}$,
A. Boccaletti$^{4}$,
\newauthor
M. Carbillet$^{5}$,
D. Mouillet$^{6}$
and I. Smith$^{5}$ \\
$^{1}$LAM, UMR 6110, CNRS, Universit\'e de Provence, 38 rue Fr\'ed\'eric Joliot-Curie, 13388 Marseille Cedex 13, France \\
$^{2}$CRAL, UMR 5574, CNRS, Universit\'e Lyon 1, 9 avenue Charles Andr\'e, 69561 Saint Genis Laval Cedex, France \\
$^{3}$CRAL, UMR 5574, CNRS, Universit\'e Lyon 1, ENS Lyon, 46 all\'ee d'Italie, 69364 Lyon Cedex 07, France \\
$^{4}$LESIA, UMR 8109, Observatoire de Paris, CNRS, Universit\'e Paris-Diderot, 5 place Jules Janssen, 92195 Meudon Cedex, France \\
$^{5}$Laboratoire Fizeau, UMR 6525, CNRS, Universit\'e de Nice Sophia Antipolis, Observatoire de la C\^ote d'Azur, Parc Valrose,\\
\hspace{2mm}06108 Nice Cedex 2, France \\
$^{6}$LAOG, UMR 5571, CNRS, Universit\'e Joseph-Fourier, BP 53, 38041 Grenoble Cedex 9, France
}

\begin{document}

\date{Accepted 2010 April 23.  Received 2010 April 21; in original form 2010 March 23.}

\pagerange{\pageref{firstpage}--\pageref{lastpage}} \pubyear{2010}

\maketitle

\label{firstpage}

\begin{abstract}
The direct detection of exoplanets has been the subject of intensive research in the recent years. Data obtained with future high-contrast imaging instruments optimized for giant planets direct detection are strongly limited by speckle noise. Specific observing strategies and data analysis methods, such as angular and spectral differential imaging, are required to attenuate the noise level and possibly detect the faint planet flux. Even though these methods are very efficient at suppressing the speckles, the photometry of the faint planets is dominated by the speckle residuals. The determination of the effective temperature and surface gravity of the detected planets from photometric measurements in different bands is then limited by the photometric error on the planet flux. In this work we investigate this photometric error and the consequences on the determination of the physical parameters of the detected planets. We perform detailed end-to-end simulation with the CAOS-based Software Package for spectro-polarimetric high-contrast exoplanet research (SPHERE) to obtain realistic data representing typical observing sequences in Y, J, H and Ks bands with a high contrast imager. The simulated data are used to measure the photometric accuracy as a function of contrast for planets detected with angular and spectral+angular differential methods. We apply this empirical accuracy to study the characterization capabilities of a high-contrast differential imager. We show that the expected photometric performances will allow the detection and characterization of exoplanets down to the Jupiter mass at angular separations of 1.0\as and 0.2\as respectively around high mass and low mass stars with 2 observations in different filter pairs. We also show that the determination of the planets physical parameters from photometric measurements in different filter pairs is essentialy limited by the error on the determination of the surface gravity.
\end{abstract}


\section{Introduction}
\label{sec:introduction}

Since the detection of the first exoplanet orbiting a main sequence star, 51~Peg, a large population of these objects has been discovered covering a wide range of masses and orbital periods mostly with indirect methods such as radial velocities measurements and transits (see e.g. \citealt{santos2008} for a review). Although mainly sensitive to planets with period of less than 10~years, radial velocities surveys have found stars which start to show long term trends indicating possible low mass companions orbiting at large orbital separations \citep{wittenmyer2007}. The wide use of adaptive optics (AO) systems and coronagraphy in large telescopes instrumentation for high-contrast imaging has allowed to start probing the vicinity of nearby stars for low mass companions at large orbital separations. Over the last decade, a handful of objects close to the planetary mass regime have been imaged with existing instruments, such as 2M~1207~b \citep{chauvin2005a}, DH~Tau~B \citep{itoh2005}, GQ~Lup~b \citep{neuhauser2005}, AB~Pic~b \citep{chauvin2005b}, CHXR~73~B \citep{luhman2006}, and more recently Fomalhaut~b \citep{kalas2008}, 1RXS~J1609~b \citep{lafreniere2008}, $\beta$~Pic~b \citep{lagrange2008} and the triple system around HR~8799 \citep{marois2008b}. However, the large uncertainty on the mass of these objects may place some of them in the sub-stellar rather than the planetary mass regime.

The next generation of planet finding instruments currently being built will combine: (i) extreme AO systems with a large number of actuators \citep{angel1994,stahl1995,langlois2001} to reach very high corrections in the near-infrared \citep{fusco2006,allercarpentier2008}; and (ii) high-efficiency coronagraphs such as the apodized pupil Lyot coronagraph (\citealt{soummer2005} and references therein) or the achromatic 4-quadrant phase mask \citep{rouan2000,mawet2006} to obtain optimal star extinction. GPI (Gemini Planet Imager) for Gemini South \citep{macintosh2006} and SPHERE (Spectro-Polarimetric High-contrast Exoplanet REsearch) for the ESO-VLT \citep{beuzit2006} are the two leading instruments of that category. These will both start operation in 2011, along with HiCIAO (High-contrast Coronagraphic Imager for Adaptive Optics) for Subaru \citep{hodapp2008}. They will aim at detecting exoplanets down to the Jupiter mass (\MJup) around nearby young stars by reaching contrast values of 15 to 17.5~mag ($10^{-6}$ to $10^{-7}$) at angular separations of $\sim$0.1\as. Both GPI and SPHERE will incorporate diffraction limited integral field spectrographs (IFS) in the near infrared, allowing to obtain images simultaneously at several wavelengths. SPHERE will also incorporate a differential imager named IRDIS (InfraRed Dual Imaging Spectrograph, \citealt{dohlen2008a}) that will provide simultaneous images at two close wavelengths in either one of its 5 different filter pairs over the Y to Ks bands. 

These instruments will allow to use different observing strategies such as Spectral Differential Imaging (SDI, \citealt{racine1999}) or Angular Differential Imaging (ADI, \citealt{marois2006a}) to obtain data which will be analyzed using advanced methods such as Spectral Deconvolution \citep{sparks2002} or Localy Optimized Combination of Images (LOCI, \citealt{lafreniere2007}) for IRDIS data. Specific methods of signal extraction have also been developed within the SPHERE consortium \citep{mugnier2008,smith2009} to be used in the data reduction pipeline of IRDIS. Data analysis methods are of extreme importance when it comes to the detection of faint objects in coronagraphic images dominated by speckles. In particular, the precise estimation of the object flux after applying these methods is critical for calibration of planetary mass objects model atmospheres (\citealt{allard2001,allard2003,allard2007}, and in preparation; \citealt{burrows2006,ackerman2001,tsuji2005}) based on effective temperature (\teff) and surface gravity (\logg), and the corresponding evolutionary models \citep{burrows1997,chabrier2000,baraffe2003,fortney2005}. The only currently available method for mass estimation, when no dynamical mass estimations are known for the object, consists in incorporating photometric measurements and their error assessment into these physical models. Although spectroscopy is the method of choice for characterization, it may not be possible to obtain high-quality spectra of extremely faint sources \citep{vigan2008,janson2010}. 

In this paper we investigate the photometric limitations in high-contrast data obtained at different wavelengths with a dual-band imager like IRDIS, and we study how it translates in terms of characterization of planetary mass objects. After reminding in Sect.~\ref{sec:limitations_high_contrast_imaging} the origin of the speckle noise which is the fundamental limitation in high-contrast imaging and the methods to overcome it, we present in Sect.~\ref{sec:end_to_end_simulations} the end-to-end simulations of IRDIS which were performed to obtain a realistic observing sequence. Section~\ref{sec:photometric_accuracy} briefly describes the detection limits obtained with ADI and SDI+ADI data analysis methods before studying the performances in terms of photometric accuracy. Finally, in Sect.~\ref{sec:photometric_characterization} we advocate a filter pair procedure for IRDIS and present an analysis of the characterizations which will be possible from aperture photometry.

\section{Limitations in high-contrast imaging}
\label{sec:limitations_high_contrast_imaging}

Detecting very faint planetary objects requires to obtain diffraction limited images with high-order AO systems in order to overcome the large contrast ratio between the star and the planet with coronagraphy. In high-Strehl ratio coronagraphic images, the factor that limits the accessible dynamical range is the speckle noise \citep{soummer2007} induced by atmospheric phase residuals and instrumental quasi-static aberrations not corrected by the AO system. The quasi-static speckles are caused by the instrumental aberrations that slowly change during a long exposure. Telescope orientation, temperature variations or rotating optical elements cause small mechanical variations in the optical elements which make the speckle pattern evolve. Long time exposures are typically decomposed in a series of short exposure images of a few seconds during which the atmospheric residuals are averaged out, forming a smooth halo over which the quasi-static speckles are superimposed, because their coherence time is much longer than the atmospheric residuals \citep{langlois1998,macintosh2005,hinkley2007}. To optimize AO performances and speckle rejection, observations can be performed in pupil stabilized mode, leading to a very high stability of the star Point Spread Function (PSF) and a slow rotation of the field of view during the observations, at a rate which depends on the star position in the sky.

The speckle noise can be reduced by subtracting a reference PSF from each science frame in order to remove the star halo and speckles, and possibly reveal a faint planetary object. This reference PSF can be obtained by observing a reference star taken in the same observing conditions (parallactic angle and atmospheric conditions if possible) as the original target to reproduce a similar pattern of quasi-static speckles. This is a very time consuming task because the time spent on the reference star is equal to the one spent on the target to precisely match both PSF, and the aberrations between the two stars cannot be exactly reproduced. The reference can also be built from the science frames by using either spectral or angular information. 

The SDI method has first been proposed by \citet{racine1999} for faint companion detection. It has been extensively studied \citep{marois2000}, and subsequently tested on sky with TRIDENT on the CFHT \citep{marois2005} and with NACO on the VLT \citep{lenzen2004}. The technique relies on the fact that planetary objects have large intrinsic molecular features in their spectrum, while the host star has a relatively flat spectrum. By taking simultaneously two images of a system at two close wavelengths located around one of these sharp features and subtracting them, the star contribution can be partially eliminated, and the planet signal revealed. SDI is most effective when used for detecting cool companions that show deep molecular absorption bands caused by H$_2$O, CH$_4$ and NH$_3$ at low \teff according to state of the art planetary mass objects atmosphere models. With carefully selected filter pairs, a contrast of several magnitudes on the planet flux between the two filters can be obtained. However, the presence of the molecular features expected for the detection of cool planetary companions should not be taken for granted since recent atmosphere models \citep{fortney2008}, as well as observations of 2M1207b \citep{mohanty2007,patience2010} and the HR8799 planets \citep{marois2008b,metchev2009,janson2010}, seem to show that non-equilibrium CO/CH$_4$ chemistry could play an important role in young low surface gravity objects. In particular, the CH$_4$ band head near 1.6~\mic could appear at much lower \teff than predicted by current atmosphere models. The SDI method is straightforward to implement: the images taken at $\lambda_1$ need to be spatially rescaled to account for the spectral dependence of the PSF and subtracted from the images at $\lambda_0$, with a possible amplitude correction factor to minimize the residual speckle noise. The main advantage of SDI is to significantly reduce the seeing halo, but it is intrinsically limited by speckle chromaticity and differential aberrations when going through two separate optical paths.

The ADI method proposed by \citet{marois2006a} requires observations made in pupil-stabilized mode. It uses the field rotation to build an optimized reference PSF that contains very little signal from the planet. For each image $I_i$, a reference PSF is calculated using images taken before and images taken after, and for which a field rotation of at least 1.5~$\lambda/D$ has occurred in between. These images are then combined to eliminate the planet signal and produce a reference PSF that is subtracted from the image $I_i$. These operations are performed for all images in annuli of increasing radius. A thorough description of the complete procedure can be found in Sect.~5.2 of \citet{marois2006a}. This technique is essentially limited by the temporal evolution of the speckles which cannot be controled. The global efficiency of the ADI method is controled by the rotation rate of the field of view, which depends on the star declination, and by the angular separation, which constrains the actual motion on the detector. At the latitude of ESO-Paranal observatory (-24\degre~3\am~38\as) and for a star at declination $\delta = -45$\degre, the field rotation varies between 0.006~deg~s$^{-1}$ at an hour angle of $\pm2$h and 0.011~deg~s$^{-1}$ at an hour angle of 0h. This defines a strong constraint on the telescope time necessary to calibrate the speckles.

Finally, the SDI and ADI methods can be efficiently combined to further reduce the speckle noise. SDI is first performed on short exposure images acquired simultaneously to remove the fast varying atmospheric residuals that have not been averaged out. ADI is subsequently applied on this set of data to combine the images with different angular positions of the field of view.

\section{End-to-end simulations}
\label{sec:end_to_end_simulations}

\begin{figure*}
  \centering
  \includegraphics[width=1.0\textwidth]{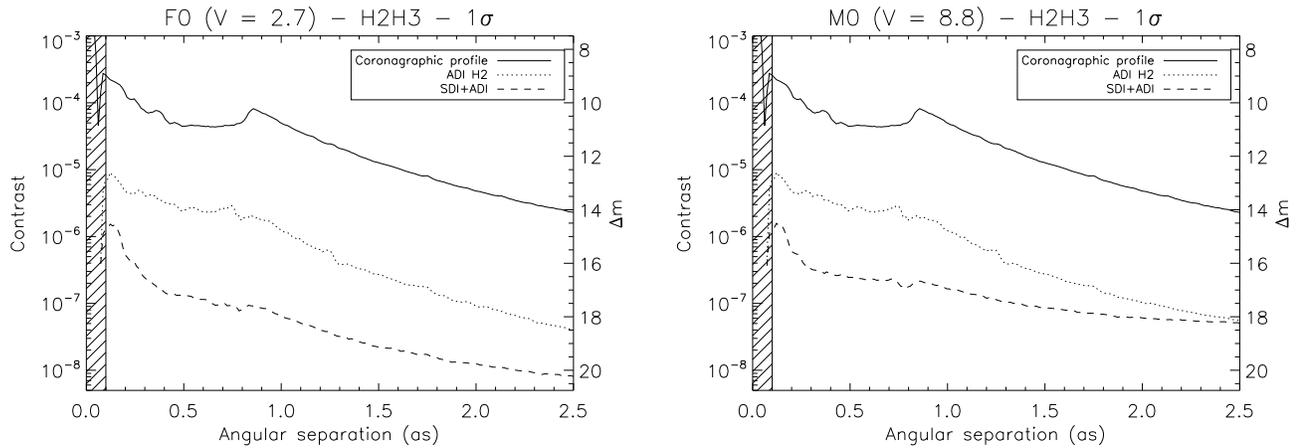}
  \caption{1-$\sigma$ noise levels after applying ADI and SDI+ADI data analysis methods in H band for high mass (left, F0 at 10~pc, V~=~2.7) and low mass (right, M0 at 10~pc, V~=~8.8) stars during a 4~h exposure time. The hatched area below 0.1\as is covered by the opaque coronagraphic mask. The coronagraphic profile is calculated by the average of the coronagraphic image in annuli of increasing radius, and the different noise levels by the standard deviation of the residuals in the same annuli. All curves are normalized to the maximum of the PSF without coronagraph.}
  \label{fig:detection_limits_H2H3}
\end{figure*}

\begin{table}
 \caption{List of IRDIS filter pairs.}
 \label{tab:irdis_filter_pairs}
 \centering
 \begin{tabular}{ccccc}
 \hline
 Pair name & \multicolumn{2}{c}{Filter 0}  & \multicolumn{2}{c}{Filter 1 } \\
           & $\lambda_0$ & $R_0$           & $\lambda_1$ & $R_1$            \\
           & (\mic)      &                 & (\mic)      &                 \\
 \hline
 Y2Y3     & 1.020       & 20              & 1.073       & 20              \\
 \hline
 J2J3     & 1.190       & 25              & 1.270       & 25              \\
 \hline
 H2H3     & 1.587       & 30              & 1.667       & 30              \\
 H3H4     & 1.667       & 30              & 1.731       & 30              \\
 \hline
 K1K2     & 2.100       & 20              & 2.244       & 20              \\
 \hline
 \end{tabular}
\end{table}

\begin{table}
 \caption{Atmosphere models included in our library.}
 \label{tab:models_library}
 \centering
 \begin{tabular}{lcc}
 \hline
 Model              & \teff       & \logg      \\
                    & (K)         & (dex)      \\
 \hline
 AMES-COND$^{\mathrm{a}}$  &  350 -- 1300 & 2.5 -- 6.0 \\
 BT-SETTL$^{\mathrm{b}}$ & 1100 -- 2300 & 4.5 -- 5.5 \\
 AMES-DUSTY$^{\mathrm{c}}$ & 1700 -- 2500 & 2.5 -- 6.0 \\
 BSH$^{\mathrm{d}}$   &  700 -- 2000 & 4.5 -- 5.5 \\
 \hline
 \end{tabular}
 \begin{list}{}{}
 \item[$^{\mathrm{a}}$] \citet{allard2003}
 \item[$^{\mathrm{b}}$] \citet{allard2007}
 \item[$^{\mathrm{c}}$] \citet{allard2001}
 \item[$^{\mathrm{d}}$] \citet{burrows2006}
 \end{list} 
\end{table}

A complete end-to-end model of SPHERE has been developed to test the instrument performances and different data analysis methods. This model is a diffractive code written in IDL (Interactive Data Language) based on the CAOS (Code for Adaptive Optics System) problem solving environment \citep{carbillet2004} with a specific package developed for the SPHERE project \citep{carbillet2008}.

Realistic data cubes have been simulated to represent typical 4 hours exposure
with IRDIS and an apodized Lyot coronagraph at different wavelengths where the star goes from -2 to +2 hour angle. Every data cube is composed of 144 images representing a cumulative 100 seconds esposure each, and several parameters are modified in the course of the simulation to take into account the variations of optical aberrations on a long timescale. The seeing and wind speed have been varied on ranges typical for the ESO-Paranal observatory, respectively $0.85 \pm 0.15\as$ and $14.2 \pm 4.6$~m~s$^{-1}$. For each individual images, the AO-corrected atmosphere was simulated by a set of 100 decorrelated phase screens. The typical millisecond timescale of the uncorrected atmospheric residuals is not considered here: we assume that on a 100-second timescale, these residuals are averaged out and produce a smooth halo; only the correlated residuals with timescales longer than a few hundred seconds will remain. From the instrumental point of view, variations of the beam shift as well as rotation of the entrance window, atmospheric dispersion corrector (ADC) and derotator have been translated into wavefront error. Chromatic shifts associated to the ADC have been calculated from its optical design. Slow achromatic drifts, such as defocus and tilt, associated with temperature changes have been added. Finally, differential aberrations between the two filters in the differential imager have been taken into account: considering the prototypes of IRDIS DBI filters \citep{dohlen2008b}, 7.55~nm~RMS of differential aberrations have been introduced.

In these end-to-end simulations, the Fresnel propagation of the wavefront is not considered. However, the overall impact of Fresnel propagation has been evaluated in separate simulations (not detailed here) where pre-coronagraphic and post-coronagraphic propagation effects have been simulated. The main result is that while the region beyond AO cut-off (0.8\as in H band) is mostly dominated by the pre-coronagraphic propagation effects, resulting in a loss of up to a factor 2 in contrast, the inner region is affected by a mix of both effects, resulting in a loss of at most 1.5.

Four complete data cubes have been simulated, corresponding to the filter pairs Y2Y3, J2J3, H2H3 and K1K2 of IRDIS (Table~\ref{tab:irdis_filter_pairs}). The output of the diffractive code is a series of normalized coronagraphic and non-coronagraphic images of the star at the two wavelengths of the considered filter pair. A second code was used to create data cubes representing realistic planetary systems. For each star, 3 series of planets separated by 120\degre have been simulated at 0.2\as, 0.5\as, 1.0\as, 1.5\as and 2.0\as, taking into account the slow field rotation which is a function of the star elevation. The star was chosen at a declination of -45\degre and an hour angle of -2~h at the beginning of the simulated observation, representing a total field rotation of $\sim$120\degre.

To calculate realistic photometry we used standard Kurucz models \citep{kurucz1979,castelli2003} for stars with spectral types regularly distributed from F0 to M0 at a distance of 10~pc (V~=~2.7 to 8.8). For the planets we constituted a library of $\sim$220 synthetic spectra including the AMES-Dusty models of \citet{allard2001}, the BT-Settl models of \citet{allard2007}, the AMES-Cond models of \citet{allard2003} and the models of \citet{burrows2006} with effective temperature ranging from \teff~=~350~K to \teff~=~2500~K, and surface gravity ranging from \logg~=~2.5 to \logg~=~6.0. The steps in the grids of models are of 100~K in \teff and 0.5 in \logg. We assume these models are complementary, and Table~\ref{tab:models_library} gives a list of the models with the \teff and \logg ranges over which they are considered. For each filter pair we generated 66 data cubes with different combination of star and planet models to cover contrast values from 5~mag to 16.5~mag ($\sim$$2 \times 10^{-7}$ to $\sim$$10^{-2}$).

In each data cube, sky contribution has been added to match typical values for ESO-Paranal observatory. Thermal background from the instrument was also included: the value is low in I, J and H bands ($<$~2~photon~sec$^{-1}$~pixel$^{-1}$), while it becomes significant in K band (60 to 220~photon~sec$^{-1}$~pixel$^{-1}$). The code also accounts for the global throughput of the instrument and the atmospheric transmission, but does not consider OH lines variability. Finally, a realistic amount of noise for the IRDIS detector was included in the images: photon noise, flat field noise (0.1\%) and read-out noise (10 e$^-$/read). 

The final output of the photometric code represents a 4~h observation with IRDIS after standard cosmetic correction and calibrations (dark, sky background and thermal background subtraction, flat field division, bad pixels correction). Due to the large number of parameters taken into account and the important computing time required for the simulation, only one data set representing a standard case has been produced. This means that all generated data cubes present the same speckle pattern, and only differ by the photometric and noise values.

ADI and SDI+ADI data analysis methods were then applied on all our simulated data cubes to attenuate the speckle noise. The ADI data analysis was implemented in IDL following the algorithm described by \citet{marois2006a}: frames separated by 2.0~$\lambda/D$ were selected and combined in 5 annuli covering our simulated planets to produce 2 final images for data taken at $\lambda_0$ and at $\lambda_1$. The SDI data analysis was implemented in IDL using a custom routine for the precise spatial rescaling (L. Mugnier, private communication) based on zero-padding in both real and Fourier spaces. ADI was then applied on the subtraction of data taken at $\lambda_0$ and at $\lambda_1$ to produce the final SDI+ADI image.

\begin{figure*}
  \centering
  \includegraphics[width=0.49\textwidth]{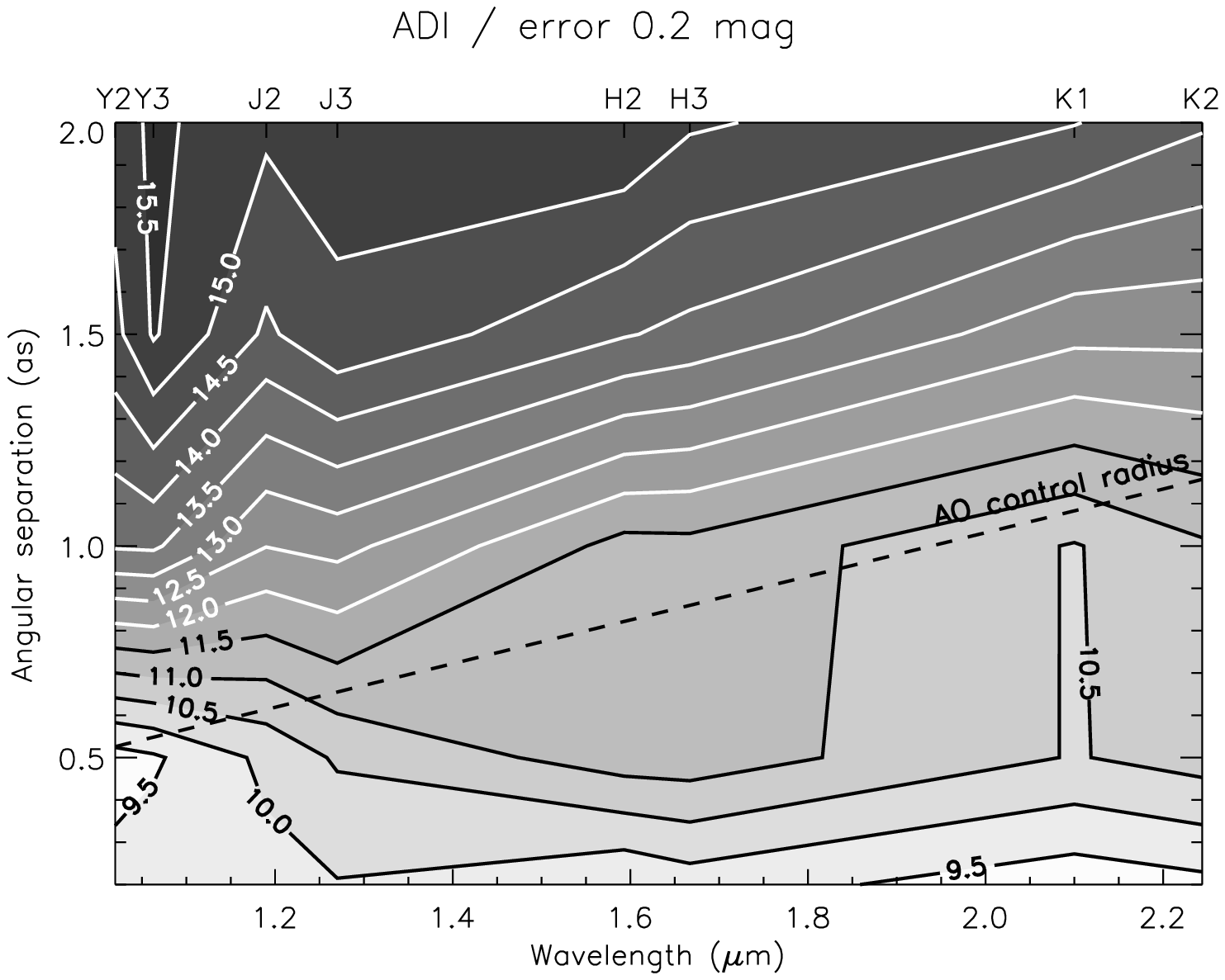}
  \hfill
  \includegraphics[width=0.49\textwidth]{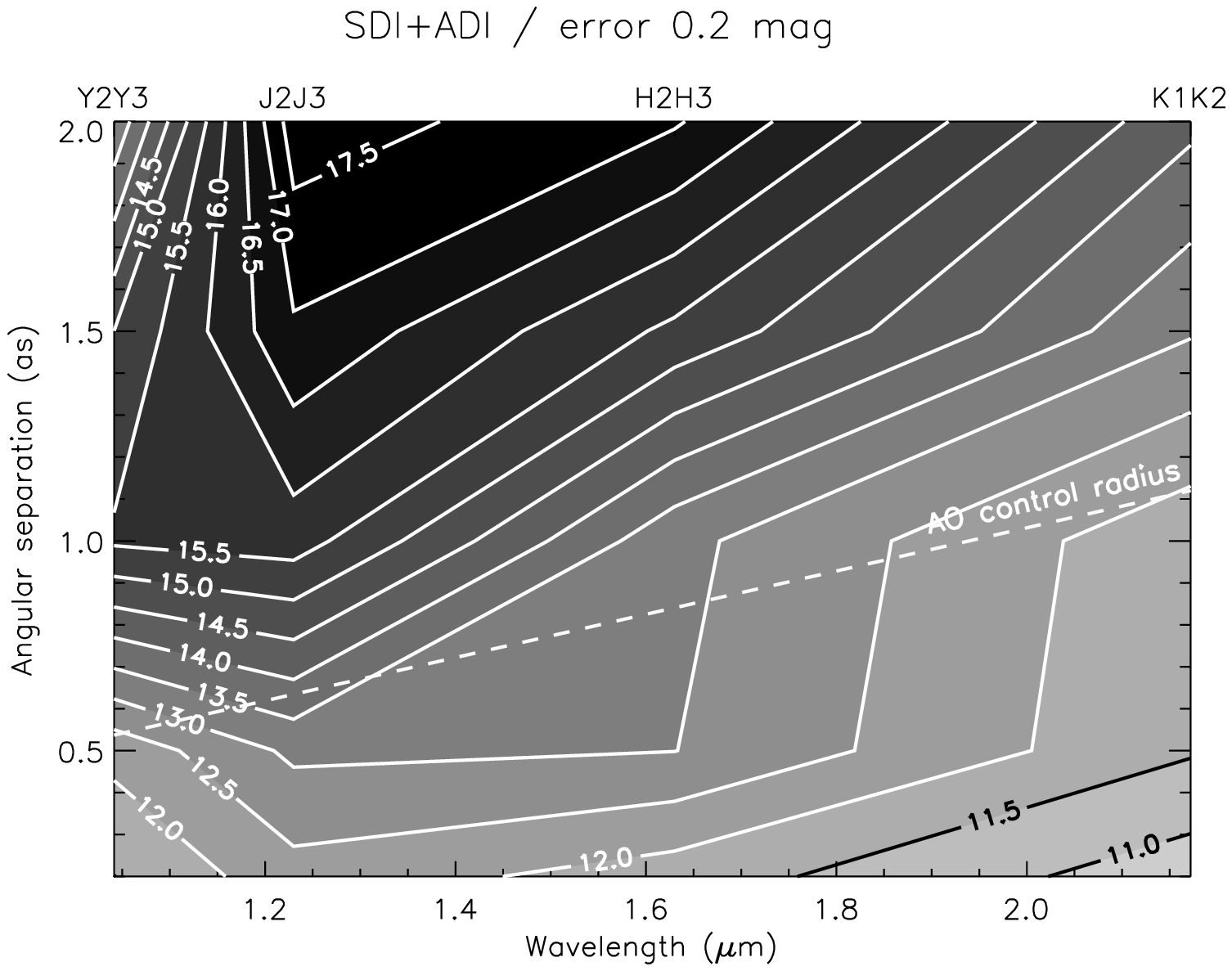}
  \caption{Magnitude difference between the star and the planet for which the photometric precision is better than 0.2~mag, as a function of wavelength and angular separation using ADI (left) and SDI+ADI (right) data analysis methods. The oblique dashed-line shows the AO control radius limit.}
  \label{fig:flux_error_summary}
\end{figure*}

\section{Photometric accuracy}
\label{sec:photometric_accuracy}

\subsection{Noise level with ADI and SDI+ADI}
\label{sec:noise_level}

1-$\sigma$ noise level were evaluated from the data products of the ADI and SDI+ADI data analysis methods by measuring the standard deviation of the residual speckle noise in annuli of increasing radius, normalized to the maximum of the PSF without coronagraph. Although it is known from \citet{goodman1968}, \citet{aime2004}, \citet{fitzgerald2006} and \citet{smith2009} that the speckle noise statistics in AO-corrected images with and without coronagraph is not Gaussian. \citet{marois2008a} have shown that residual noise after applying ADI on $\sim$20 or more images is quasi-Gaussian. ADI was applied on our data cubes with more than 20 images in every case, so we will consider the residual noise as Gaussian and we will use standard 5-$\sigma$ level for detection limits

Fig.~\ref{fig:detection_limits_H2H3} shows the 1-$\sigma$ noise level for H2H3 filter pair in two different regimes: a high flux case corresponding to a high mass star (F0 star at 10~pc, V~=~2.7) and a low flux case corresponding to a lower mass star (M0 star at 10~pc, V~=~8.8). The ADI noise level in filter H3 is not shown because it is at the same level as in filter H2. In high flux, the speckle noise attenuation is almost constant from 0.1\as (inner-working angle of the coronagraph) to 2.5\as, both with ADI and SDI+ADI, allowing to reach a contrast of $\sim$16~mag at 0.2\as and more than 20~mag at 2.5\as. In low flux, the level of the background noise (sky and instrumental thermal emission, read-out noise) becomes limiting, and the attenuation reaches an almost constant level in SDI+ADI at 1.5\as. The noise levels are similar in Y, J and H bands, but in K band, where the thermal emission is 10 to 15 times higher than in J or H band, the accessible contrast in low flux regime is limited at 15.5~mag.

If we compare these 1-$\sigma$ noise level to the ones derived for IRDIS Long Slit Spectroscopy (LSS) mode \citep{vigan2008}, we see that they are comparable between ADI and LSS. However, SDI+ADI clearly brings an improvement of 2 to 3 magnitudes compared to LSS, justifying the need to quantify characterization capabilities in DBI mode for planets that are not detectable with LSS.

\subsection{Planet flux estimation}
\label{sec:planet_flux_estilation}

We have estimated the signal of all planets detected at more than 5-$\sigma$ with aperture photometry in a 2.44~$\lambda/D$ radius aperture. The aperture is calculated to take into account the effect of using a round aperture on square pixels. The measured signal has been converted to a flux in phot~s$^{-1}$~m$^{-2}$ with the formula:

\begin{equation}
  \label{eq:flux_norm}
  f_i = \frac{S_i}{E_e~S_{\mathrm{Tel}}~T_r~t_i},
\end{equation}

\noindent where $S_i$ is the measured signal on image $i$, $E_e$ is the encircled energy in the aperture, $S_{\mathrm{Tel}}$ is the telescope collecting surface, $T_r$ is the transmission of the atmosphere, telescope and instrument, and $t_i$ is the exposure time for image $i$ (100~s in our case).

We consider that we are in a case where we know the value of the encircled energy $E_e$ in the aperture. This value varies mostly with seeing conditions because the AO correction will concentrate more energy in the PSF core when seeing improves. Moreover, we consider that the planet position is known exactly to center the aperture on the planet PSF and avoid photometric error bias induced by inaccurate centering. Finally we also take into account the error induced by the field rotation. When the field rotates, it will slowly smear the planet PSF, especially at large angular separations. The effect in our case is significant at separations larger than 1.0\as because we simulated long exposures for individual images (100~s). In practice exposures for individual images will typically last a few seconds to avoid detector saturation, reducing the effect of PSF smearing up to a few arcseconds.

\subsection{Photometric accuracy in ADI}
\label{sec:photometric_accuracy_adi}

For our simulated test case, the planet flux has been evaluated in all data cubes after using ADI data analysis method. In each filter pair and for each simulated planet, 2 independent values are obtained at $\lambda_0$ and $\lambda_1$. They are compared to the flux value introduced at the beginning of the simulation to evaluate the photometric error. Figure~\ref{fig:flux_error_summary}, left, illustrates the photometric performance as a function of wavelength and angular separation. The contours indicate the contrast value between the star and planet below which the photometric precision is better than 0.2~mag. Such a precision is necessary in order to be able to disentangle between different planet masses when comparing actual measurements to evolution nary models. We see two major effects: (i) the photometric performance clearly depends on wavelength, and (ii) there are two different regimes depending on the position compared to the AO control radius. The first effect is directly related to the chromaticity of the PSF: in speckle-limited regime the noise attenuation is almost constant with angular separation compared to the coronagraphic profile, and the level of the coronagraphic profile linearly depends on wavelength. The second effect is related to the AO correction inside the control radius. Inside that region we see a stabilization of the performance: 0.2~mag photometric precision can be reached up to contrast of 10 to 11~mag ($10^{-4}$ to $4 \times 10^{-4}$) from 0.2\as to the AO control radius, which extends from 0.5\as in Y band to 1.0\as in K band. Outside of the AO control radius, the photometric performance increases almost linearly with angular separation at all wavelengths to reach contrast values of 14 to 15~mag ($2.5 \times 10^{-6}$ to $10^{-6}$) around 2.0\as. These numbers are given in the context of our simulated test case, but the general effects should be similar for any data obtained with high contrast coronagraphic imagers.

\subsection{Photometric accuracy in SDI+ADI}
\label{sec:photometric_accuracy_sdi_adi}

Similarly to the noise level, using the SDI+ADI data analysis method improves the photometric accuracy. However, using SDI+ADI will only provide an estimation of the differential flux of the planet between the 2 filters, contrary to ADI which provides an absolute measurement. To preserve the planet differential flux, the amplitude correction factor usually applied for SDI in the subtraction is taken equal to a fixed value of 1. The photometric error estimated with SDI+ADI follows the same variations as for ADI, but at higher contrast values. Figure~\ref{fig:flux_error_summary}, right, illustrates the photometric performance as a function of wavelength and angular separation in SDI+ADI. The trends are similar to ADI alone, but the chromatic effect is less significant because the PSF chromaticity has been mitigated by the SDI part of the analysis. Compared to ADI alone, the contrast values at which a 0.2~mag photometric error is reached are 1.5 to 2.5~mag higher. At shorter wavelengths, in Y2Y3 filters, performances at separations larger than 1.0\as decrease. This effect is related to the size of the aperture for photometry which is very small in Y band (4~pixels in diameter), and to the field rotation which has a strong effect on encircled energy at separations larger than 1.0\as in Y band. Considering shorter exposures for individual images where the field rotation is negligible would decrease the photometric errors in that particular case.

\subsection{Empirical photometric accuracy}
\label{sec:empirical_photometric_accuracy}

\begin{figure*}
  \centering
  \includegraphics[width=1.0\textwidth]{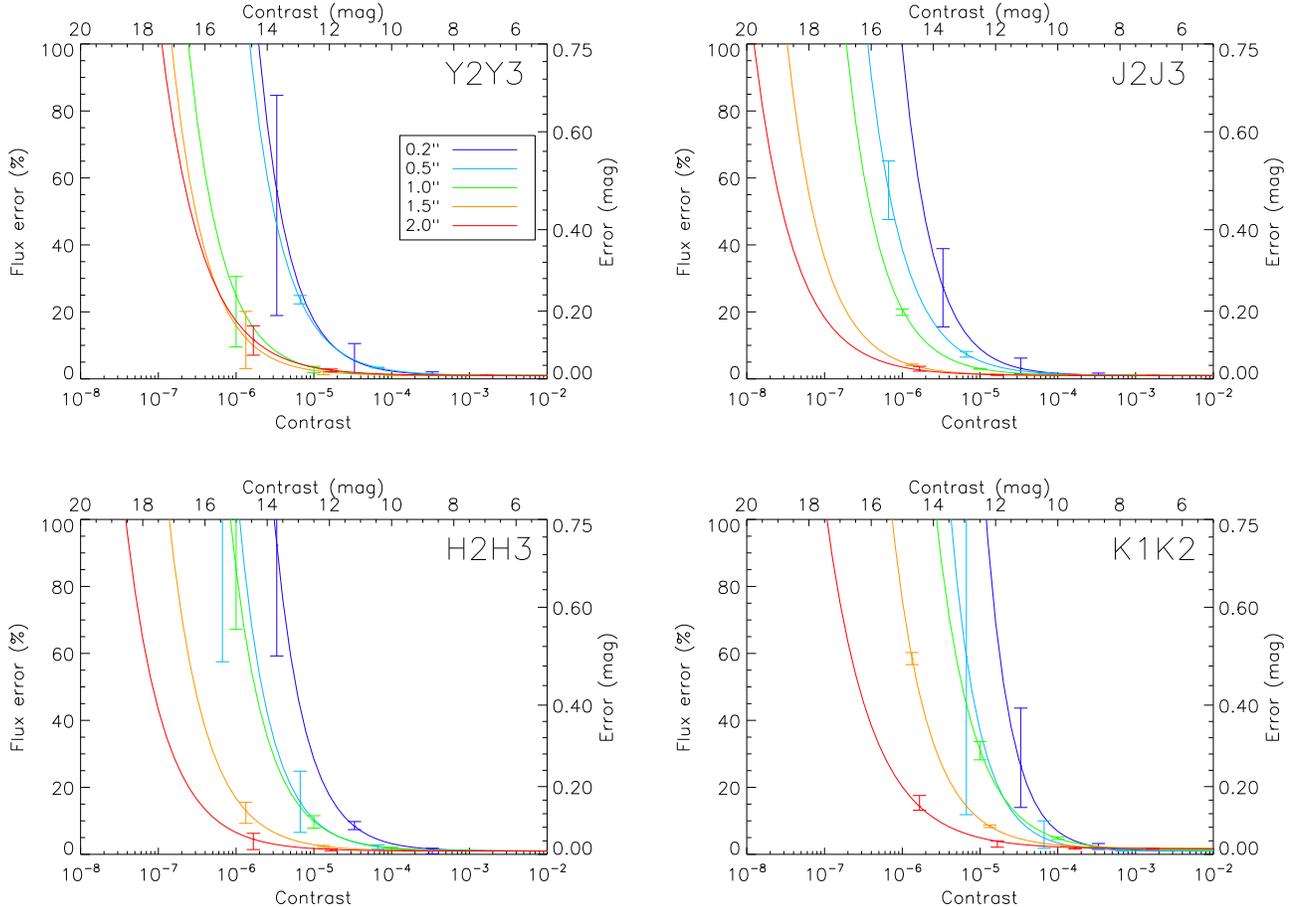}
  \caption{Empirical photometric error as a function of contrast in IRDIS filter pairs using either ADI or SDI+ADI data analysis methods. Errors bars have been represented only for a small set of data points. Their amplitude is defined by the optimal and pessimistic error curves described in the text.}
  \label{fig:error_curves_phot_all}
\end{figure*}

We hereafter combine the photometric accuracy obtained in ADI and SDI+ADI to define empirical photometric error curves for each filter pair as a function of contrast. The photometric error curves as a function of contrast at each angular separation have been fitted with the empirically defined function:

\begin{equation}
  \label{eq:empirical_phot_err}
  \mathrm{phot_{err}} = \frac{p_1}{c^{p_2}}+p_3
\end{equation} 

\noindent where $\mathrm{phot_{err}}$ is the photometric error, $c$ the contrast and $(p_1, p_2, p_3)$ the fitted parameters. This function approaches the measured points with a precision of $\sim$1\%. The fitting has been performed for ADI and SDI+ADI. To take into account the scattering of the error with the planet position in the images, different cases have been considered at each separation, corresponding to the 3 different simulated planet positions: a standard case with an average photometric error, an optimal case corresponding to the lowest estimation of the error and a pessimistic case corresponding to the upper estimation of the error. These empirical photometric errors are plotted in Fig.~\ref{fig:error_curves_phot_all} for the 4 simulated filter pairs. The amplitude of the error bars is defined by the optimal and pessimistic error curves described above. We assume that the photometric error in ADI is the same in the two filters of a pair, which is legitimate given the amplitude of the error bars. These empirical error curves lie in the same range as the expected photometric accuracy of other data analysis methods developed within the SPHERE consortium by \citet{mugnier2008} and \citet{smith2009}.

\begin{table}
 \caption{Contrast limit over which the differential photometric error in SDI+ADI becomes smaller than the photometric error in ADI.}
 \label{tab:methods_limits}
 \centering
 \begin{tabular}{ccccc}
 \hline
Separation & \multicolumn{4}{c}{Filter pair} \\
           &  Y2Y3  & J2J3  & H2H3  & K1K2    \\
(arcsec)   &  (mag) & (mag) & (mag) & (mag)   \\
 \hline
 0.2\as    &    6.8 &   6.8 &   6.5 &   6.8   \\
 0.5\as    &    6.0 &   6.8 &   8.3 &   8.0   \\
 1.0\as    &   10.8 &   9.3 &   8.0 &   7.2   \\
 1.5\as    &   12.8 &  11.0 &  11.0 &  10.0   \\
 2.0\as    &        &  11.5 &  12.2 &  12.2   \\
 \hline
 \end{tabular}
\end{table}

Table~\ref{tab:methods_limits} gives for each filter pair and each angular separation the contrast value at which the photometric error in ADI becomes lower than the differential photometric error in SDI+ADI. These values give the contrast at which it becomes more interesting in terms of photometric error to obtain a differential flux estimation. As explained in Sect.~\ref{sec:photometric_accuracy_sdi_adi}, aperture photometry in Y band is extremely sensitive to errors introduced by the position of the aperture or the field rotation because the aperture is very small. This is why in Y2Y3 pair at 2.0\as there is no contrast limit between ADI and SDI+ADI: for that particular case the flux estimation error is slightly better in ADI than SDI+ADI.

\section{Photometric characterization}
\label{sec:photometric_characterization}

In this section we evaluate characterization capabilities of IRDIS in imaging mode, i.e. how well the physical parameters \teff and \logg of the planets can be estimated from photometric measurements in different spectral bands.

\subsection{Characterization simulation}
\label{sec:characterization_simulation}

To estimate the characterization capabilities of IRDIS, we performed a new simulation using as input the 5-$\sigma$ detection limits obtained from Sect.~\ref{sec:noise_level} and the empirical error curves obtained in Sect.~\ref{sec:empirical_photometric_accuracy}. The goal of the simulation was to test the efficiency of all filter pair sequences for characterization at different stellar magnitudes and for a large number of planetary atmosphere models. These simulations are based on current state of the art atmosphere models. Although these models will clearly evolve with new detections in the future, they allow to test the expected performances of IRDIS, as well as to estimate the intrinsic errors of our signal extraction and comparison to models. It has been performed for all stellar types and atmosphere models included in our library (see Table~\ref{tab:models_library}).

For the simulation we assume that a same planetary system is observed with different filter pairs in a given order. For each possible combination of parameters (filter pairs sequence; star magnitude; angular separation; planet atmosphere model) we proceed as follow: the star and planet fluxes are calculated in the filters of the first pair; if the planet is not detectable (considering the 5-$\sigma$ detection limit), simulation for that combination of parameters is stopped; if it is indeed detectable, a photometric measurement is obtained. Depending on the contrast between the planet and the star in each filter, different informations are obtained: 2 direct photometric measurements if the planet is detectable with ADI in both filters, a differential flux measurement if the planet is only detectable with SDI+ADI or a direct and a differential measurement if the planet is only detectable in ADI in one of the filters. Once the flux measurements are obtained, the photometric error is determined from the empirical error curves and added to the measured values to obtain lower and upper limits to the planet flux. Models that can correspond to these limits are then searched in our models library. If only one model corresponds, we stop iterations considering that the planet has been fully characterized within the limits of the grid of atmosphere models that is used. If many models match those limits, we switch to the next filter pair in the sequence and the same process is started again. In a given sequence, each filter will bring some additional information that will help to find the appropriate atmosphere model and constrain the values of \teff and \logg. At the end of the filter pairs sequence, 4 distinct outcomes are possible:

\begin{itemize}
\item \emph{Non-Detection (ND):} the planet is not detected in the first filter pair; the sequence is stopped. \\

\item \emph{Non-unique Characterization (NC):} the planet is detected at least in the first filter pair of the sequence; at the end of the sequence many models match the flux measurements and they have different values of \teff and \logg. \\

\item \emph{\teff Characterization (TC):} the planet is detected at least in the first filter pair of the sequence; at the end of the sequence many models match the flux measurements and they all share the same value for \teff but not for \logg. \\

\item \emph{Full Characterization (FC):} the planet is detected at least in the first filter pair of the sequence; at the end of the sequence only one model matches the flux measurements, which means that the \teff and \logg values have been determined.
\end{itemize}

The TC and FC are considered within the limits of the atmosphere models grid, which is 100~K in \teff and 0.5 in \logg, i.e. the \emph{full characterization} corresponds to the determination of \teff and \logg with an error equal to the limits of the models grid. Another important point is that in practice the photometric error will have to be estimated from the science data itself. Although, we consider here an ideal case where the photometric error is known, other data analysis methods such as the one proposed by \citet{smith2009} will allow direct estimation of the error from the data, with a precision that it compatible with the results presented here.

\subsection{Filter pair sequence analysis}
\label{sec:filter_pair_sequence_analysis}

\begin{table*}
 \centering
 \caption{Analysis of the filter pairs sequences.}
 \label{tab:sequences_std}
 \begin{tabular}{c|ccccc|c|c|c|c|c}
\hline
Spectral type & P1$^{\mathrm{a}}$ & P2$^{\mathrm{a}}$ & P3$^{\mathrm{a}}$ & P4$^{\mathrm{a}}$ & P5$^{\mathrm{a}}$ & ND$^{\mathrm{b}}$ & NC$^{\mathrm{c}}$ & TC$^{\mathrm{d}}$ & FC$^{\mathrm{e}}$ & TC+FC \\
   &      &      &      &      &      & (\%) & (\%)        & (\%)         & (\%)          & (\%)  \\
\hline
M0 & H2H3 &      &      &      &      &  3 & $20_{-1}^{+1}$ & 0            & $77_{-1}^{+1}$ & $77_{-1}^{+1}$\rule{0cm}{0.4cm} \\
V = 8.8   & H2H3 & Y2Y3 &      &      &      &  3 & $ 2_{-1}^{+1}$ & 0            & $95_{-1}^{+1}$ & $95_{-1}^{+1}$\rule{0cm}{0.35cm} \\
H = 5.3   & H2H3 & Y2Y3 & J2J3 &      &      &  3 & $ 1_{-1}^{+0}$ & 0            & $96_{-1}^{+1}$ & $96_{-0}^{+1}$\rule{0cm}{0.35cm} \\
          & H2H3 & Y2Y3 & J2J3 & H3H4 &      &  3 & $ 1_{-1}^{+0}$ & 0            & $96_{-1}^{+1}$ & $96_{-0}^{+1}$\rule{0cm}{0.35cm} \\
          & H2H3 & Y2Y3 & J2J3 & H3H4 & K1K2 &  3 & $ 1_{-1}^{+0}$ & 0            & $96_{-1}^{+1}$ & $96_{-0}^{+1}$\rule{0cm}{0.35cm} \\
\hline
K0 & H2H3 &      &      &      &      &  6 & $35_{-8}^{+1}$ & 0            & $59_{-1}^{+8}$ & $59_{-1}^{+8}$\rule{0cm}{0.4cm} \\
V = 5.9   & H2H3 & Y2Y3 &      &      &      &  6 & $ 4_{-1}^{+2}$ & $2_{-1}^{+1}$ & $88_{-2}^{+3}$ & $90_{-1}^{+2}$\rule{0cm}{0.35cm} \\
H = 4.0   & H2H3 & Y2Y3 & J2J3 &      &      &  6 & $ 2_{-1}^{+1}$ & $1_{-1}^{+0}$ & $92_{-1}^{+1}$ & $92_{-1}^{+1}$\rule{0cm}{0.35cm} \\
          & H2H3 & Y2Y3 & J2J3 & H3H4 &      &  6 & $ 2_{-1}^{+1}$ & $1_{-0}^{+0}$ & $92_{-1}^{+1}$ & $92_{-1}^{+1}$\rule{0cm}{0.35cm} \\
          & H2H3 & Y2Y3 & J2J3 & H3H4 & K1K2 &  6 & $ 2_{-1}^{+1}$ & $1_{-0}^{+0}$ & $92_{-1}^{+1}$ & $92_{-1}^{+1}$\rule{0cm}{0.35cm} \\
\hline
G0 & H2H3 &      &      &      &      &  7 & $43_{-7}^{+3}$ & 0            & $50_{-3}^{+7}$ & $50_{-3}^{+7}$\rule{0cm}{0.4cm} \\
V = 4.4   & H2H3 & Y2Y3 &      &      &      &  7 & $ 7_{-2}^{+4}$ & $3_{-1}^{+1}$ & $83_{-3}^{+5}$ & $86_{-2}^{+4}$\rule{0cm}{0.35cm} \\
H = 3.0   & H2H3 & Y2Y3 & J2J3 &      &      &  7 & $ 3_{-1}^{+1}$ & $2_{-1}^{+1}$ & $88_{-2}^{+2}$ & $90_{-1}^{+1}$\rule{0cm}{0.35cm} \\
          & H2H3 & Y2Y3 & J2J3 & H3H4 &      &  7 & $ 3_{-1}^{+1}$ & $1_{-1}^{+1}$ & $88_{-2}^{+2}$ & $90_{-1}^{+1}$\rule{0cm}{0.35cm} \\
          & H2H3 & Y2Y3 & J2J3 & H3H4 & K1K2 &  7 & $ 3_{-1}^{+1}$ & $1_{-1}^{+1}$ & $88_{-2}^{+2}$ & $90_{-1}^{+1}$\rule{0cm}{0.35cm} \\
\hline
F0 & H2H3 &      &      &      &      & 10 & $55_{-7}^{+3}$ & 0            & $35_{-3}^{+7}$ & $35_{-3}^{+7}$\rule{0cm}{0.4cm} \\
V = 2.7   & H2H3 & J2J3 &      &      &      & 10 & $11_{-2}^{+2}$ & $3_{-1}^{+0}$ & $76_{-3}^{+2}$ & $79_{-2}^{+2}$\rule{0cm}{0.35cm} \\
H = 1.5   & H2H3 & J2J3 & Y2Y3 &      &      & 10 & $ 7_{-3}^{+3}$ & $3_{-1}^{+1}$ & $80_{-4}^{+4}$ & $83_{-3}^{+3}$\rule{0cm}{0.35cm} \\
          & H2H3 & J2J3 & Y2Y3 & H3H4 &      & 10 & $ 6_{-2}^{+2}$ & $2_{-1}^{+1}$ & $82_{-3}^{+4}$ & $84_{-2}^{+2}$\rule{0cm}{0.35cm} \\
          & H2H3 & J2J3 & Y2Y3 & H3H4 & K1K2 & 10 & $ 6_{-2}^{+2}$ & $2_{-1}^{+1}$ & $82_{-3}^{+4}$ & $84_{-2}^{+2}$\rule{0cm}{0.35cm} \\
\hline
\end{tabular}
 \begin{list}{}{}
 \item[$^{\mathrm{a}}$] P1 to P5 designate the filter pairs
 \item[$^{\mathrm{b}}$] Not detected
 \item[$^{\mathrm{c}}$] No characterization
 \item[$^{\mathrm{d}}$] \teff characterization
 \item[$^{\mathrm{e}}$] Full characterization
 \end{list}
\end{table*}

The output of this simulation allows to determine the most significant filter pairs sequence for characterization, i.e. the sequence which maximizes the number of characterizations. The strategy is to progressively build an \emph{optimal sequence} by adding each time the filter pair that increases the most the number of characterizations. All possible filter pair sequences have been systematically tested to find the one that maximizes the number of \teff characterizations and full characterizations (TC+FC), as a function of stellar magnitude. The simulation shows the following important results (see Table~\ref{tab:sequences_std} for details):

\begin{itemize}
\item H2H3 is the filter pair which minimizes the number of non-detections, reflecting the fact that CH$_4$ absorption band near 1.6~\mic in cool substellar objects spectra is the optimal feature for their detection. However it should be reminded that this specific spectral feature might not always be present in lower mass objects as we mentionned in Sect.~\ref{sec:limitations_high_contrast_imaging}. \\

\item When adding a new filter pair to a sequence, adding Y2Y3 or J2J3 will increase the number of characterized models more than H3H4 or K1K2. This result does not depend on the stellar magnitude. \\

\item The scattering of the flux error between the pessimistic optimal photometric error curves has no major influence on the number of characterized models when more than one filter pair is used. In particular, we see the error bars of the different flux cases do not significantly overlap, confirming that the chosen sequence is appropriate for all cases of flux.
\end{itemize}

\begin{figure}
  \centering
  \includegraphics[width=0.5\textwidth]{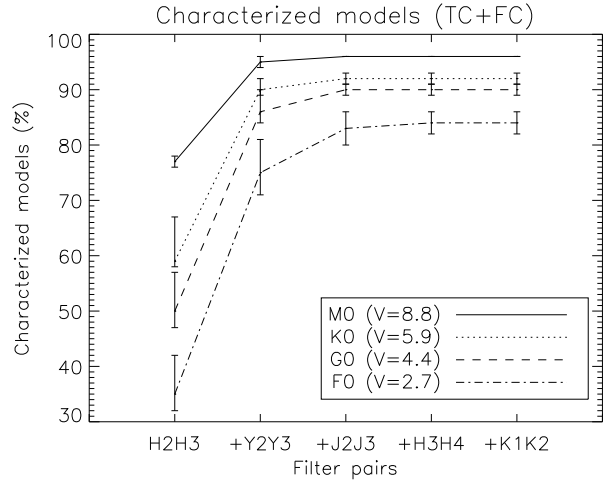}
  \caption{Proportion of models in our library characterized by adding filter pairs from the optimal filter pair sequence for 4 stellar magnitudes. The error bars are given by the optimal and pessimistic photometric error curves.}
  \label{fig:filter_pair_sequences}
\end{figure}

This last result is particularly important as it advocates for a given sequence of filter pairs and allows to set general priorities on the filter pairs for characterization. When there are no \emph{a priori} assumptions on the nature of the objects, the priorities are defined from highest to lowest as follow:

\begin{enumerate}
\item[0.] H2H3 \\
\item[1.] Y2Y3~/~J2J3 \\
\item[2.] H3H4~/~K1K2
\end{enumerate}

\noindent The H2H3~--~Y2Y3~--~J2J3~--~H3H4~--~K1K2 will be referred to as the \emph{optimal sequence} from now on, and we will only consider the standard empirical errors since the differences with the other error curves are small. Assuming the use of the optimal sequence, Fig.~\ref{fig:filter_pair_sequences} represents the proportion of characterized models (TC+FC) from our library when new filter pairs are added. The trends are identical for the 4 different stellar magnitudes. When using only H2H3 the proportion of characterized models is comprised between 30\ and 80\%, and the error bars are of $\sim$10\% for bright stars. Adding a second filter pair greatly improves the proportion of characterized models which is above 70\% for all stellar magnitudes. Adding more filter pairs confirms this trend and tends to reduce the error bars to less than 5\% for all magnitudes. The main conclusion is that most of the information for characterizing any given model is obtained using 2 filter pairs around low mass stars when the contrast is favorable, and 3 filter pairs around high mass stars for which the contrast is more challenging.

\subsection[Lowest estimations of T$_{eff}$]{Lowest estimations of \teff}
\label{sec:lowest_estimations_teff}

\begin{figure}
 \centering
 \includegraphics[width=0.5\textwidth]{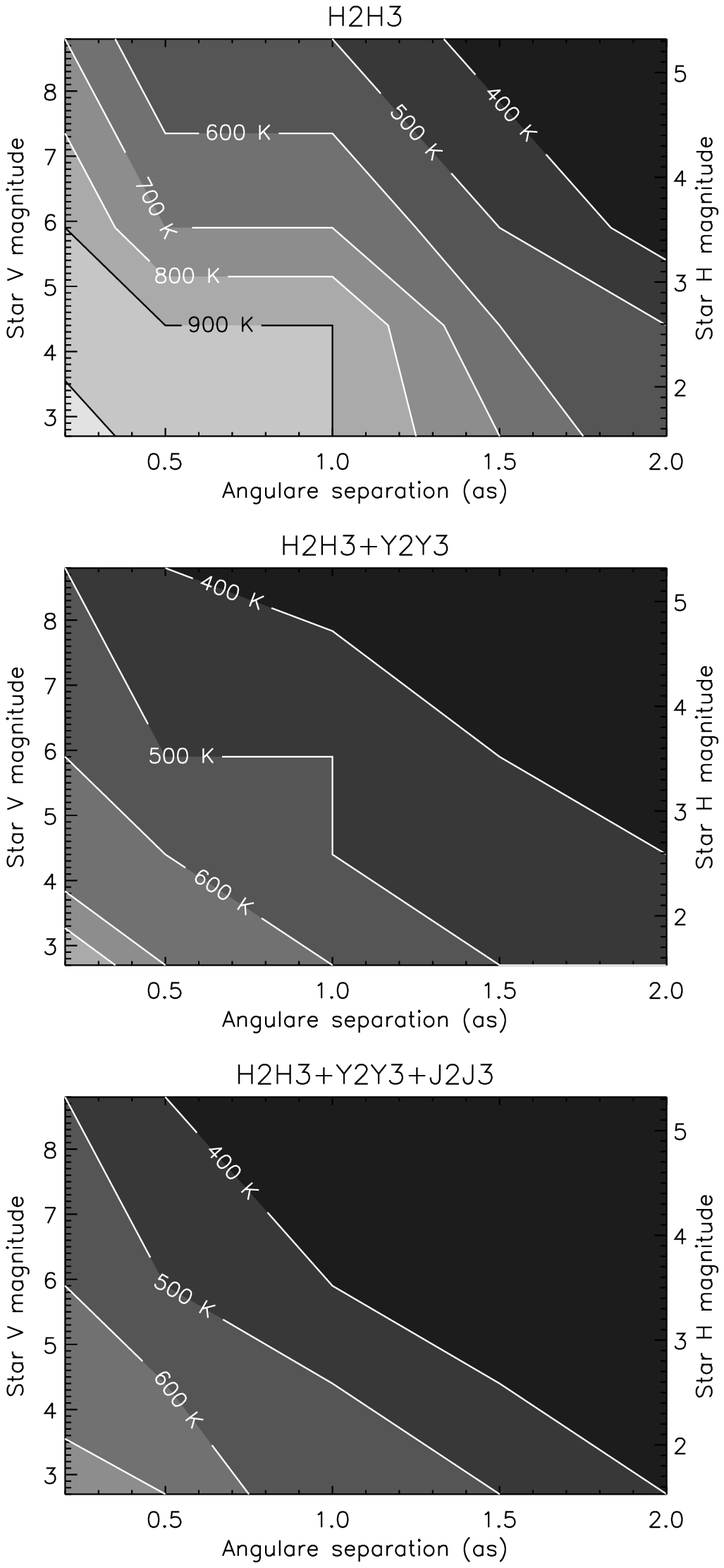}
 \caption{Smallest value of \teff which can be characterized as a function of
star magnitude (y-axis) and angular separation (x-axis) when using 1, 2 or 3
filters from the optimal sequence.}
 \label{fig:best_sequence_teff_limit}
\end{figure}

In the previous section we set priorities for characterization on the different filter pairs of IRDIS. We will now detail the lowest values of \teff that IRDIS will be able to characterize as a function of stellar magnitude and angular separation. Figure~\ref{fig:best_sequence_teff_limit} gives the smallest values of \teff which have been characterized when using 1 to 3 filter pairs from the optimal sequence. Colder planets can be detected, but we were not able to find the appropriate values of \teff and \logg. When using only H2H3, planets with \teff down to 900~K should be characterized at an angular separation of 0.2\as from high mass bright stars and 700~K from lower mass stars. Adding a second filter pair considerably improves these results by 200~K, while adding a third pair confirms these limiting values.

With the considered data analysis methods and according to the evolutionary models from \citet{baraffe2003} for the COND atmosphere models, we can estimate that in a very young system of 10~Myr, we should be able to characterize a planet of 1~\MJup with H2H3 at separations larger than 0.5\as around a low mass star (M0 at 10~pc) where the star-planet contrast is favorable, but only further than 2.0\as around a high mass star (F0 at 10~pc) where the contrast difference is larger. With two filter pairs, the limit would be 0.2\as around a low mass star and 1.0\as around a high mass star. For older systems, only planets of a few masses of Jupiter could be characterized. At 100~Myr, a Jupiter mass planet would remain out of reach for characterization with H2H3 filters around a high mass star, and only at separations larger than 1.5\as around a low mass star. At this age, the \teff limits of 700~K and 500~K which can be reached at small angular separation around high and low mass stars would respectively correspond to planets with masses of $\sim$6.5~\MJup and $\sim$3~\MJup. Using improved signal extraction methods providing more accurate photometry of the companion would certainly push down those limits.

\subsection{Study of the non-unique characterizations}
\label{sec:study_non_unique_characterizations}

\begin{figure*}
  \centering
  \includegraphics[width=1.0\textwidth]{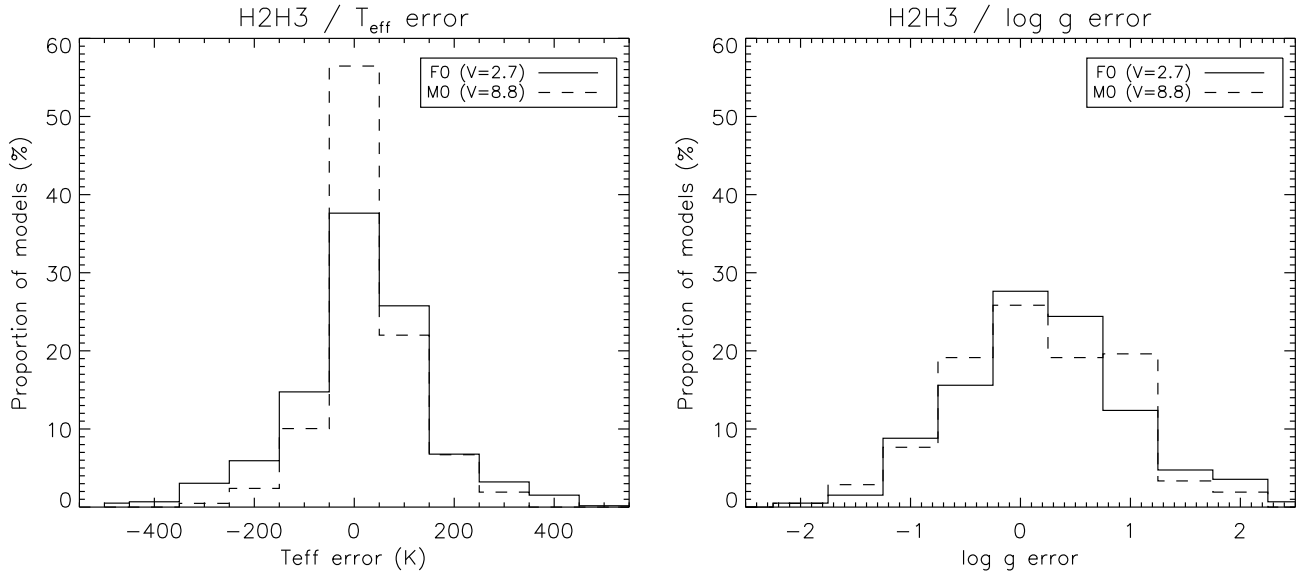}
  \caption{Distribution of the errors on \teff and \logg for non-unique characterizations with H2H3 filter pair around a high mass star (F0 at 10~pc, plain line) and a low mass star (M0 at 10~pc, dashed line) cases.}
  \label{fig:best_sequence_nochar}
\end{figure*}

NCs are the cases where several models correspond to the flux measurements in all filter pairs with which they are detected. From these remaining models, it is possible to determine if a combination of (\teff; \logg) is more represented than others, making this combination the most probable values of \teff and \logg. If several combinations are counted an equal number of times, an average value and an error can be determined for the values of \teff and \logg. In any case, the error is at least equal to the steps in the grids of models. The estimation of the most probable values for \teff and \logg has been performed for all non-uniquely characterized models at all simulated angular separations and magnitudes.

Figure~\ref{fig:best_sequence_nochar} shows an histogram of the errors on \teff and \logg when using H2H3 filter pair for high mass (F0 at 10~pc) and low mass (M0 at 10~pc) stars. NCs are mostly dominated by errors on the determination of \logg. In particular we see that around a low mass star where the contrast is more favorable, the proportion of models with no error on \teff increases by $\sim$20\%, while the errors on \logg keep the same distribution. Adding other filter pairs improves the determination of \teff for the non-unique charaterizations, in particular around high mass (brighter) stars, reaching more than 95\% for all stellar magnitudes. The determination of \logg is also improved, but even when using 3 filter pairs the number of cases where the error is less or equal to 0.5 never reaches more than 85\% around a high mass star or 95\% around a low mass star.

\subsection{Impact of errors on \teff and \logg}
\label{sec:impact_error_teff_logg}

\begin{figure}
  \centering
  \includegraphics[width=0.5\textwidth]{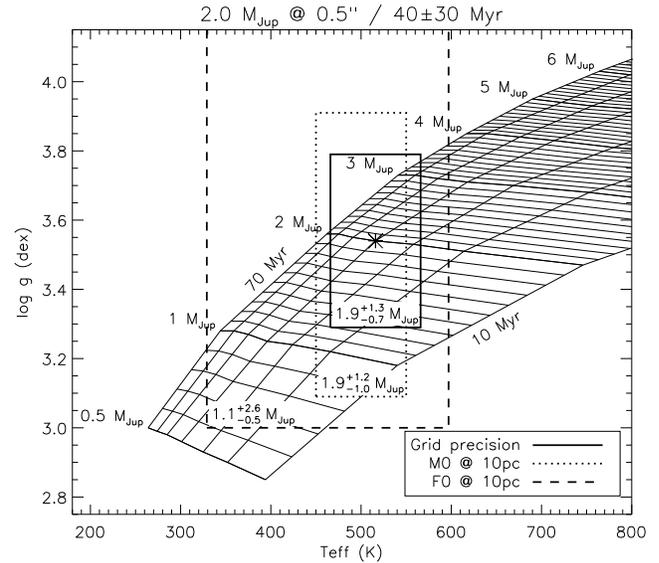}
  \caption{Isochrones for the COND planetary atmosphere models covering an age of $40 \pm 30$~Myr used for the determination of the mass of hypothetical 2~\MJup planets orbiting at 5~A.U. from M0 and F0 stars at 10~pc. The error boxes defined by the possible values for \teff and \logg of both planets are respectively represented by dotted and dashed rectangles. The planet mass derived from the models and error box is displayed at the bottom left corner of each error box (see text for an explanation on how the planet masses are derived). The position of the planet predicted by the evolutionary models is represented by the star symbol, and the error box defined by the atmosphere models grid precision is given for reference around that position by a plain rectangle.}
  \label{fig:char_error_impact}
\end{figure}

The influence of errors on the determination of \teff and \logg on the determination of the planet mass can be studied using evolutionnary models such as those published by \citet{baraffe2003} for the COND atmosphere models. The preliminary version of the SPHERE target list (S. Desidera, private communication) was used to define a standard young test case based on age considerations. The average age for targets younger than 100~Myr is $44 \pm 20$~Myr and the average error on the determation of the target age is 30~Myr. Using these values we can define two test cases considering a planet of 2~\MJup aged of $44 \pm 30$~Myr orbiting at 5~A.U. from M0 and F0 stars at 10~pc. According to the evolutionary models from \citet{baraffe2003}, such planets should have \teff~=~516~K and \logg~=~3.54~dex, resulting in a contrast of 11.9~mag and 15.6~mag in H band respectively around the M0 and F0 stars. Considering the results from Sect.~\ref{sec:lowest_estimations_teff} and \ref{sec:study_non_unique_characterizations}, the planet around the F0 star cannot be characterized with IRDIS using one filter pair, while the planet around the M0 star is close to the measured limit.

The expected spectra of these planets were introduced in our simulation to test the accuracy of extracting flux information and inversely deriving physical parameters, and to estimate their mass using evolutionary models. The results are presented in Fig.~\ref{fig:char_error_impact}: the areas covered by the values of \teff and \logg are shown as rectangular boxes on the predicted isochrones for both planetary systems. The bin size of the atmosphere model grid is also plotted for reference. For each case, the planet mass is estimated by selecting all the isochrones of the masses that cross the error box and weighting them by the integral of the isochrone inside the box. The isochrone that has the largest intersection with the error box is supposed to be the most likely (or the average if several isochrones have the same integral). The upper and lower limits of the mass are given by the highest and lowest mass isochrones that cross the box.

Around a low mass star, the parameters \teff and \logg are estimated with an accuracy close to the one given by the atmosphere model grid ($1.9^{+1.3}_{-0.7}$~\MJup), leading to an estimation of $1.9^{+1.2}_{-1.0}$~\MJup. Around a high mass star, the planet is very close to the detection limit at 0.5\as, resulting in a poor estimation of both \teff and \logg: the important photometric error in H2H3 leads to a very large uncertainty on \logg ($4.33 \pm 1.23$). The mass of the planet is then estimated to $1.1^{+2.6}_{-0.5}$~\MJup. In this case, the planet mass is stronly underestimated, and the large uncertainties on \teff and \logg lead to a large upper limit for the mass estimation. The origin of the large offset on the determination of \logg is still uncertain, and further simulations are still needed to investigate thoroughly the complete parameter space. In particular we see in this case that the age uncertainty unquestionably increases the uncertainty on the mass estimation by increasing the number of isochrones crossing the possible values of \teff and \logg. Such simulations would greatly benefit from updated homogeneous atmosphere models grids covering a large span of \teff, \logg and age.

\section{Conclusions}
\label{sec:conclusions}

Next generation instruments for ground based exoplanet direct imaging such as SPHERE and GPI will provide data intrinsically limited by the speckle noise. This noise needs to be attenuated using \emph{a posteriori} data analysis methods, such as spectral and angular differential imaging. In this paper, we have quantified the exoplanet characterization capabilities of IRDIS, the differential imager of SPHERE, using photometric and differential photometric information.

The photometric performances have been evaluated with aperture photometry on the detectable planets as a function of contrast and wavelength for a standard test case. In particular we have shown that the photometric performance strongly depends on wavelength because of the PSF chromaticity and on the position with respect to the AO control radius. With ADI, a photometric accuracy of 0.2~mag is reached inside the AO control radius for contrast values of 10 to 11~mag between the star and planet, while at larger radius the precision can be reached for contrasts up to 15~mag. With SDI+ADI, the overall photometric performance is improved, increasing by 1.5 to 2.5~mag the contrast at which a 0.2~mag precision is reached. 

We have defined empirical photometric accuracies for IRDIS in its different filter pairs, which were used to test the characterization capabilities of IRDIS with all possible combinations of filter pairs. Priorities for characterization have been set on the different filter pairs by finding the pairs that maximize the number of possible characterizations in various conditions. We showed that when there is no \emph{a priori} knowledge on the planet, the filter pairs Y2Y3 and J2J3 allow a larger number of characterizations than the pairs H3H4 and K1K2. Then we showed that using filter pair H2H3, it will be possible to characterize planets with \teff~$\simeq$~900~K around high mass stars at small angular separation and \teff~$\simeq$~700~K around lower mass stars. Adding Y2Y3 and J2J3 filter pairs allows to decrease the characterizable \teff by 200~K at all separations and for all stellar magnitudes, while considerably decreasing the number of non-characterizations for warmer planets.

Finally, we showed that non-unique characterizations, i.e. planets for which the \teff and \logg values could not be determined exactly (within the limits of the grid of atmosphere models) are mostly dominated by errors on the determination of \logg. In particular, we showed that around low mass stars, where the contrast is more favorable, the determination of \teff is largely improved, while the errors on \logg remain identical around a high mass star. Considering evolutionary models, and including typical ages from the
future SPHERE target list, we showed that such errors on the determination of a low mass planet (2~\MJup) results in a large uncertainty around a high mass bright star, but is very close to the limits fixed by the models grid around a fainter low mass star.

With this work we showed that IRDIS, the dual-band imager of SPHERE, should be able to fulfill the goal set for a high-contrast imager, that is the ability to detect and characterize planetary companions down to the Jupiter mass around nearby young stars. Similar developments could also be performed for an Integral Field Spectrograph in the future, allowing to quantify precisely the performances of SPHERE in the near-infrared, and to work on the aspect of characterization strategy for the detected objects.

\section*{Acknowledgments}

We wish to thank Silvano Desidera (INAF) for generously providing a preliminary version of the SPHERE target list with all associated physical parameters, and Laurent Mugnier (ONERA) for providing an optimized procedure for spatial rescaling of the images. 

SPHERE is an instrument designed and built by a consortium consisting of LAOG, MPIA, LAM, LESIA, Laboratoire Fizeau, INAF, Observatoire de Gen\`eve, ETH, NOVA, ONERA and ASTRON in collaboration with ESO.

%
%
\bibliographystyle{mn2e}
\bibliography{paper}

\label{lastpage}

\end{document}